\documentclass[aps,pra,twocolumn,showpacs,groupedaddress]{revtex4}
\usepackage{graphicx}
\usepackage{dcolumn}

\begin{document}

\title{Quantum Multiple Scattering: Eigenmode Expansion and Its Applications to Proximity Resonance}

\author{Sheng Li}
\email{li2@fas.harvard.edu}
\affiliation{Department of Chemistry and Chemical Biology, Harvard University, Cambridge, MA 02138}

\author{Eric J. Heller}
\email{heller@physics.harvard.edu}
\affiliation{Department of Chemistry and Chemical Biology and Department of Physics, Harvard University, Cambridge, MA 02138}

\date{\today}

\begin{abstract}

We show that for a general system of $N$ $s$-wave point scatterers, there
are always $N$ eigenmodes. These eigenmodes or eigenchannels play the same
role as spherical harmonics for a spherically symmetric
target---they give a phase shift only. In other words, the $T$~matrix of
the system is of rank~$N$ and the eigenmodes are eigenvectors
corresponding to non-$0$ eigenvalues of the $T$~matrix. The eigenmode
expansion approach can give insight to the total scattering cross
section; the position, width, and superradiant or subradiant nature of
resonance peaks; the unsymmetric Fano lineshape of sharp proximity
resonance peaks based on the high energy tail of a broad band; and other
properties. Off-resonant eigenmodes for identical proximate scatterers are approximately angular momentum eigenstates.
\end{abstract}


\maketitle

\section{Introduction}

Scattering of waves from a group of scatterers is rich in interesting
physical phenomena. Examples are electrons or phonons scattering from
defects or impurities in crystals and light scattering from conjugated
molecules or biological complexes.

Multiple scattering effects are nontrivial, especially when scatterers are
placed close, i.e., inside each other's effective radius
$\sqrt{\sigma/\pi}$, where $\sigma$ is the total cross section of a single
scatterer. If the scatterers are resonant, their effective radius can be
much larger than their physical size or force range. A classical
example is the case of weak, fixed frequency sound incident on two
proximate, small identical air bubbles in water~\cite{tol1,tol2,feu}. 
Quantum mechanically, for two scatterers placed together well within the
resonant wave length of scattered particle, extremely narrow proximity
resonance can appear~\cite{hel1,hel2}.  Three or more proximate scatterers
lead to related effects. 

In this paper we have formulated a novel general method, eigenmode
expasion, for quantum
scattering of a system of $s$-wave point scatterers of any number and
geometric configuration. The eigenmode expansion is based on Green function and is
equivalent to solving Lippmann-Schwinger equation exactly---taking all
orders of multiple scattering into account. 

It is obvious from eigenmode expansion approach that the $T$~matrix of
such a system is of rank~$N$ and we are able to construct the $T$~matrix
explicitly. Many features of the spectrum (total scattering cross section
as a function of energy) of identical proximate scatterers can be
explained simply by this approach. For example, we can naturally explain
the position, width, and superradiant or subradiant nature of resonance
peaks, and the unsymmetric Fano lineshape of sharp proximity resonance
peaks based on the high energy tail of a broad band. We are also surprised
to find off-resonant eigenmodes of {\textit{randomly distributed}}
identical proximate scatterers exhibit great symmetry and regularity.
Similar systems of $N$ interacting resonances have been previously studied,
e.g.~\cite{zel1,zel2}. Though our eigenmode expansion approach is
of course equivalent to standard scattering theory, it gives new physical insights, as we demonstrate below.

In addition, the quantum scattering of $N$ proximate $s$-wave scatterers is related to
the superradiance problem~\cite{dic, gro}. Again, our eigenmode expansion gives
new insights to superradiance.

\section{\label{sec1} \label{sec2} Review of Quantum Multiple Scattering}


We start by introducing our model for $s$-wave point scatterers. Consider a single elastic scatterer placed at $\vec{r}=0$,  The asymptotic form
of the total wave function is:
\begin{equation}
\label{fundamental} \psi_{\vec{k}}(\vec{r})  \cong e^{i \vec{k} \cdot
\vec{r}} + f(\vec{k},\hat{r}) \frac{e^{i k r}}{r} \mbox{ for } r \to
\infty.
\end{equation}
The first term on the RHS is the incoming plane wave, and the second term
is the scattered wave. The
angular dependent scattering amplitude $f(\vec{k},\hat{r})$
satisfies the optical theorem:
$
4 \pi/k\mbox{ Im}f(\vec{k},\hat{r}=\hat{k})=\sigma(\vec{k}),
$
where the total cross section 
$ \sigma(\vec{k})\equiv\int|f(\vec{k},\hat{r})|^2\,d\hat{r}. $
The $T$~matrix of the scatterer in the subspace of energy shell
$|\vec{k}|=k$, $T_k$, can be defined as an operator acting on functions of
$\hat{r}$: 
\[
T_k y(\hat{r}) \equiv \frac{k}{4\pi} \int
f(k\hat{r}',\hat{r})y(\hat{r}')\, d\hat{r}'.
\]
We assume the force range of the scatterer is small compared to
the wavelength of the scattered particle, so it only scatters the
$s$-wave component ($j_0(kr)$) of the incoming wave. The now angular independent scattering amplitude is related to the phase angle $\delta(k)$ by $f(k)=\frac{1}{k}\sin \delta(k) e^{i \delta(k)}$. The $T$ matrix is of rank 1. $T_k$ has only one non-0 eigenvalue 
$\sin \delta(k) e^{i \delta(k)}$, whose corresponding eigenvector is the 0th order sperical harmonic. 

A simple scatterer with one internal Breit-Wigner type resonance can be modeled as:  
\begin{equation} 
\label{f} 
\label{bw}
f(k)=-\frac{1}{k}\frac{\gamma_0}
{\frac{k^2}{2}-\frac{k_0^2}{2}+i\gamma_0}. 
\end{equation} 
The total cross section has a Lorentzian shape (dashed line in
Fig.~\ref{fig1}a): 
\[
\sigma(E)=\frac{4\pi}{k^2}\frac{\gamma_0^2}{(E-E_0)^2+{\gamma_0^2}},
\]
where $E=k^2/2$ and $E_0=k_0^2/2$. Resonance corresponds to the pole of
$f(E)$ on complex $E$ plane, at $E_0-i\gamma_0$,
where the ``scattered wave'' can exist without the incoming wave.
Physically $E_0$ and $\gamma_0$ correspond to position and half width of the
resonance peak. 

We further assume the force range of the scatterer is so small that it can be considered as existing only at one point. In this point scatterer model, Eq.~(\ref{fundamental}) holds
not only asymtotically, but over all space.
 Since any
incoming wave $\phi(\vec{r})$ of energy $E=k^2/2$ can be written as a
superposition of plane waves with
$|\vec{k}|=k$, the total wave function can be written as:
\begin{equation}
\label{gpoint}
\psi(\vec{r})=\phi(\vec{r})+\phi(0)f(k)G_k(r),
\end{equation}
where the free space Green function $G_k(r)=e^{ikr}/r$. 
The only significant difference is the extra factor
$\phi(0)$, the amplitude of the incoming wave at the point scatterer. 

We are now ready to tackle the multiple-scatterer problem~\cite{her}. The total wave function for a system of $N$ scatterers fixed at positions
$\vec{r}_1$, $\vec{r}_2$, $\cdots$, $\vec{r}_N$ is: 
\begin{equation}
\label{phisc}
\psi(\vec{r})=\phi(\vec{r})+\sum_{i=1}^{N}\psi_i(\vec{r_i})f_iG_i(\vec{r}),
\end{equation}
where $f_i$ is the scattering amplitude of the $i$-th scatterer,
$G_i(\vec{r})$ is the free space Green function from $\vec{r}_i$: 
$
G_i(\vec{r})=e^{ik|\vec{r}-\vec{r}_i|}/{|\vec{r}-\vec{r}_i|},
$
and for simplicity we have omitted all implicit $k$ (energy)  dependence. The $\phi(0)$ in Eq.~(\ref{gpoint}) is replaced by the
amplitude at the $i$-th scatterer of the $i$-th effective incoming wave
$\psi_i(\vec{r})$---the sum of the incoming wave and waves scattered by
all scatterers except by the $i$-th scatterer itself:
\begin{equation}
\label{nbasic}
\psi_i(\vec{r})=\phi(\vec{r})+\sum_{j\not=i}\psi_j(\vec{r_j})f_jG_j(\vec{r}).
\end{equation}
The above equation provides a system of linear
equations for $\psi_i(\vec{r}_i)$. We can express them explicitly in
matrix form. Define vector for incoming wave $\vec{\phi}$ and vector for
effective incoming waves $\vec{\psi}$ as:
$
\phi_i \equiv \phi(\vec{r}_i)$, $ \psi_i \equiv \psi_i(\vec{r}_i)$ ($i=1,\cdots,N$). 

Define the $N{\times}N$ free space Green matrix $\mathbf{G}$ as: 
\[
G_{ij}\equiv \left\{ \begin{array}{ll}
    G_j(\vec{r}_i) & \mbox{for $i\not=j$,}\\
    0 & \mbox{for $i=j$,} \end{array} 
\right. 
\]
and matrix $\mathbf{F}$ as: 
\[
\mathbf{F} \equiv diag\{f_1, f_2, \cdots, f_N\}.
\] 
Substitute $\vec{r}$ with $\vec{r}_i$ in Eq.~(\ref{nbasic}), we get
\begin{equation}
\label{l_s}
\vec{\phi}=\mathbf{M}\vec{\psi}, \mbox{ or
  }\vec{\psi}=\mathbf{M}^{-1}\vec{\phi},
\end{equation}
where
\[
\mathbf{M}\equiv \mathbf{1}-\mathbf{GF}. 
\]
The physical implication of Eq.~(\ref{l_s}) is clear. Formally
\[
\mathbf{M}^{-1}=\mathbf{1}+\mathbf{GF}+(\mathbf{GF})^2+\cdots,
\]
each term $(\mathbf{GF})^l$ $(l=0,1,\cdots)$ means the incoming wave is
scattered $l$ times, since $\mathbf{F}$ means scattering once by one
scatterer, and $\mathbf{G}$ means free propagation from one scatterer to
another. An integral equation similar to Eq.~(\ref{l_s}) is known in general
scattering theory as the Lippmann-Schwinger equation~\cite{lip}. Resonances occur at poles
of $\mathbf{M}^{-1}(E)$ on complex $E$ plane. 

The scattered wave $\phi_{sc}(\vec{r})$ depends only on $\vec{\phi}$, the
amplitudes of the incoming wave at scatterers.  Asymptotically
\begin{equation}
\label{x1}
\phi_{sc} \cong (\sum_{i=1}^N \psi_i f_i
e^{-ik\hat{r}\cdot\vec{r}_i})\frac{e^{ikr}}{r} \mbox{ for }r\to\infty. 
\end{equation}
If the incoming wave is a plane wave $e^{i\vec{k}\cdot\vec{r}}$, the sum
in parenthesis of Eq.~(\ref{x1}) is nothing else but the
scattering amplitude $f(\vec{k}, \hat{r})$.

\section{Formulation of Eigenmode Expansion}

For a single scatterer, for incoming wave $j_0(kr)$
($=({G(r)-G^*(r)})/{2ik}$), the scattered wave is proportional to
$G(r)$---the outward part of the incoming wave---with a phase factor. Is there any incoming wave for multiple scatterers such that the scattered wave is proportional to the outward part of the incoming wave with a phase factor? This is
equivalent to looking for eigenvectors of the $T$~matrix of the whole system. 
In the spherically symmetric case, diagonalization of $T$~matrix is essentially spherical harmonic
partial wave expansion. Some textbooks have discussed the formal
diagonalization of the $T$~(or $S$)~matrix for a general system, and its
application to scattering of particles with spin and scattering
reactions \footnote{See, e.g., M. L. Goldberger and K. M. Watson,
\textit{Collision Theory} (John Wiley \& Sons, Inc., 1964), pp346--352 and
pp372--376.}. We have found, as will be shown immediately, that
for a system of $s$-wave point scatterers of any number and geometric
configuration, the $T$~matrix can be analytically represented under some
special basis, and can be diagonalized easily.

From Eq.~(\ref{nbasic}), we see the
scattered wave of multiple scatterers is a
superposition of $G_i(\vec{r})$'s, so we try to write the incoming wave as
a superposition of $j_0(k|\vec{r}-\vec{r_i}|)$'s:
\begin{equation}
\label{x5}
\phi(\vec{r})=\sum_{i=1}^{N} q_{ij} j_0(k|\vec{r}-\vec{r_i}|)=\sum_{i=1}^{N}
q_{ij}\frac{G_i(\vec{r})-G_i^*(\vec{r})}{2ik},
\end{equation}
or $\vec{\phi}=\mathbf{J}\vec{q_j}$, where $\vec{q_j}=(q_{1j},\cdots,q_{Nj})^T$, and the $N \times N$ spherical Bessel
matrix $\mathbf{J}$ is defined as: 
\[
J_{ij} \equiv j_0(k|\vec{r_i}-\vec{r_j}|).
\]
We have
$\vec{\psi}=\mathbf{M}^{-1}\vec{\phi}$,
and from Eq.~(\ref{phisc}), we get
\[
\phi_{sc}(\vec{r})=\sum_{i=1}^{N}G_i(\vec{r}) 
[\mathbf{FM}^{-1}\mathbf{J}\vec{q_j}]_i =\frac{1}{k} \sum_{i=1}^{N}
G_i(\vec{r}) [\mathbf{TJ}\vec{q_j}]_i,
\]
where matrix $\mathbf{T}$ is defined as
\[
\mathbf{T} 
\equiv k \mathbf{FM}^{-1}.
\]
If $\vec{q_j}$ is an eigenvector of $\mathbf{TJ}$: 
\[
\mathbf{TJ}\vec{q_j}=\lambda_j\vec{q_j}\mbox { ($j=1,\cdots,N$)},
\]
the scattered wave
\[
\phi_{sc}(\vec{r}) = \frac{\lambda_j}{k} \sum_{i=1}^{N}
G_i(\vec{r})q_{ij} 
\]
is proportional to the outward part of the incoming wave. The angular dependence of the incoming and scattered wave is
\begin{equation}
\label{a6} 
\label{b5} 
\label{basis}
y_j(\hat{r}) \equiv \sum_{i=1}^{N} q_{ij} e^{-ik\hat{r}\cdot\vec{r}_i}.
\end{equation} 
 Each $\vec{q_j}$ or $y_j(\hat{r})$ can be regarded as an
{\textit{eigenmode}} or eigenchannel of the system. $q_{ij}$ is the relative amplitude of the
scattered wave from the $i$-th scatterer. Eigenvalue $\lambda_j$ is related to the phase
angle of the $j$-th eigenmode $\Delta_j$ by
\[
\lambda_j=\sin{\Delta_j}e^{i\Delta_j}. 
\]

Eigenmodes have some basic properties that can be proved straight forwardly. (1) $\vec{q_j}$ can always be chosen real due to time reversal
symmetry:
\[
q_{ij}^*=q_{ij}.
\]
(2) The eigenmodes obey orthonormal relations upon normalization: 
\[
{\vec{q_i}}^T\mathbf{J}\vec{q_j}=\frac{1}{4\pi} \delta_{ij}, \mbox{ or }
\int y_i^*(\hat{r})y_j(\hat{r})\, d\hat{r}=\delta_{ij}. 
\]
(3) $\lambda_j$'s and $y_j(\hat{r})$'s are actually eigenvalues and
eigenvectors of the $T$~matrix:
\[
Ty_j(\hat{r})=\sin \Delta_j e^{i\Delta_j} y_j(\hat{r}). 
\]
It is clear now that $y_j(\hat{r})$'s here play the same role
as spherical harmonics in spherically symmetric case. Spherical harmonics
are eigenchannels under special conditions, namely, when the system is
spherically symmetric.

Any incoming wave can be written as:  $\phi(\vec{r})=\sum_{i=1}^{N} c_i
j_0(k|\vec{r}-\vec{r}_i|)+\phi'(\vec{r})$, where $\phi'(\vec{r})$ is
orthogonal to each $j_0(k|\vec{r}-\vec{r}_i|)$. Immediately we see the
$\phi'(\vec{r})$ part does not scatter at all, because each scatterer only
scatters $s$-wave and $\phi'(\vec{r})$ is orthogonal to each
$j_0(k|\vec{r}-\vec{r}_i|)$. For a
single scatterer, the $T$~matrix is of rank $1$. For a
system of $N$ scatterers, the $T$~matrix is of rank~$N$, since the system
only scatters incoming waves belonging to the $N$-dimensional function
space the sum in Eq.~(\ref{x5}) spans. 

There is a simpler explanation for the rank of the $T$~matrix. Recall from
section~\ref{sec2} that the scattered wave $\phi_{sc}(\vec{r})$ depends only on
$\vec{\phi}$, amplitudes of the incoming wave at scatterers. For all the infinitely degenerate
incoming waves of energy $E=k^2/2$, we can have a basis in which all but
$N$ waves have $0$~amplitude at all $\vec{r}_i$'s. Consequently, the
system only scatters the $N$ waves which have non-$0$ amplitudes at
$\vec{r}_i$'s, and does not scatter the rest at all.

In eigenmodes $y_j(\hat{r})$ (plus all other linearly independent
functions of $\hat{r}$ orthogonal to all $y_j(\hat{r})$'s) basis, the
$T$~matrix has a diagonal form:
$
T=diag\{ \sin\Delta_1 e^{i\Delta_1}, \cdots, 
  \sin\Delta_N e^{i\Delta_N}, 0, \cdots\}.
$
The eigenmodes are actually eigenvectors corresponding to non-$0$
eigenvalues of the $T$~matrix. In fact, matrix $\mathbf{T}$ is
representation of $T$~matrix in the (nonorthogonal)  $b_j(\hat{r})  \equiv
\frac{1}{\sqrt{4 \pi}} \sum_{i=1}^{N} J_{ij} e^{-i k \hat{r} \cdot
\vec{r}_i}$ basis (or $\mathbf{JTJ}$ is representation of $T$~matrix in
the $\frac{1}{\sqrt{4\pi}} e^{-i k \hat{r} \cdot \vec{r}_j}$ basis). This
can be verified if you view Eq.~(\ref{basis}) as transformation of
basis. 

We can use eigenmode expansion to calculate the total cross section for an
incoming plane wave. Optical theorem can also be verified. The total scattering cross section not only depends on
the energy (or $k$), but also the direction $\hat{k}$ of the incoming wave
. This is what we would
expect for a system without spherical symmetry. The average of total
cross section over all possible incoming directions for a given
energy is independent of $\hat{k}$. From now on, we will use ``total cross section'' in this sense. The
total cross section
\[
\sigma=\frac{4\pi}{k^2}\sum_{j=1}^{N}\sin^2\Delta_j
\]
is a sum of contributions from each eigenmode~\footnote{The average of total cross section over all possible
incoming directions can also be obtained from methods described in section~\ref{sec2}. See ref.~\cite{her}.}.

Numerical evidence shows that for scatterers with the properties of
Eq.~(\ref{bw}), each phase angle $\Delta_j$ monotonically increases
from $0$ to $\pi$ as the energy goes from $-\infty$ to $\infty$. 
Therefore each eigenmode will contribute a resonance peak
to total cross section (if the peaks do not overlap). The position of the
peak is at where $\Delta_j=\pi/2$, and the width is determined by how fast
$\Delta_j$ passes through $\pi/2$ and nearby region as a function of
energy. For a system of $N$ scatterers, there will be $N$ resonance peaks
in total cross section, whose positions and widths are consistent with the
$N$ poles of $\mathbf{M}^{-1}(E)$ on complex $E$ plane.

\section{Applications of Eigenmode Expansion to Proximity Resonance}

\begin{figure}
\includegraphics{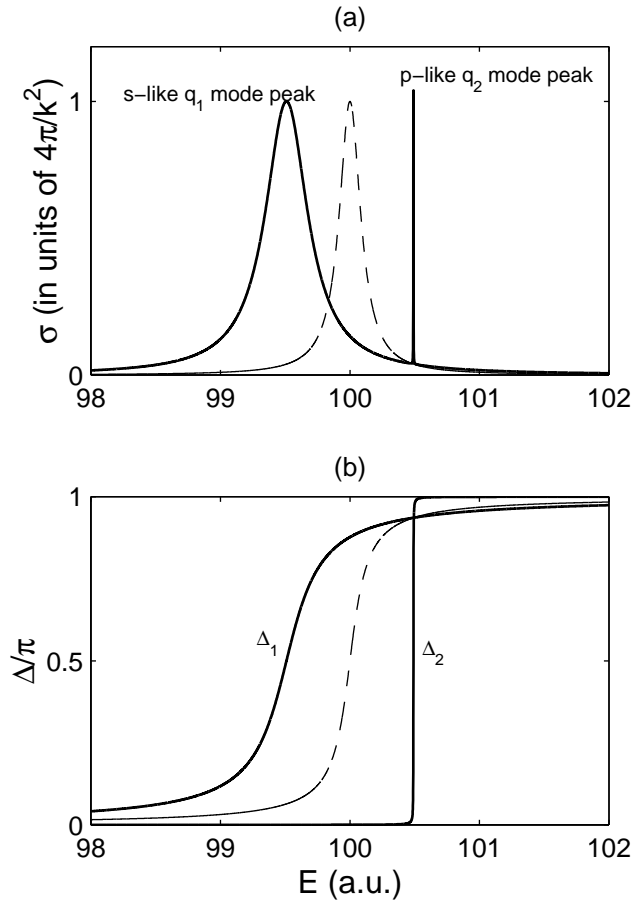}
\caption{\label{fig1} Solid lines: (a) total cross section $\sigma$ and (b) phase
angles $\Delta_j$ of each eigenmode
as a function of energy of a system of two identical scatterers
separated by distance $d=0.2/k_0$. Each
scatterer is specified by Eq.~(\ref{bw}) with $E_0=100(\mbox{a.u.})$ and $\gamma_0=0.1$. For comparison, the total cross section and phase angle
of a single scatterer are also shown (dashed lines).}
\end{figure}

First we consider the simplest case of multiple scattering: two identical
scatterers separated by distance $d$. It is easy to verify that
$\mathbf{F}$, $\mathbf{M}$, and $\mathbf{J}$ all have the form of $\left(
\begin{array}{cc} a&b\\b&a \end{array}\right)$, so does
$\mathbf{TJ}=k\mathbf{FM}^{-1}\mathbf{J}$. Thus without knowing $f(k)$
explicitly, we can immediately say the two (unnormalized) eigenmodes are: 
\begin{equation}
\label{2modes}
\vec{q_1}=\left(\begin{array}{d}1\\1\end{array}\right) 
\mbox{ and }
\vec{q_2}=\left(\begin{array}{d}1\\-1\end{array}\right),
\end{equation}
corresponding to the two scatterers oscillating perfectly in and out of
phase with the same strength. Eq.~(\ref{2modes}) is valid at
any energy. Fig.~\ref{fig1} shows an example of the total cross section and phase
angles as a function of energy. For $kd<1$, i.e., the scatterers are
placed within each other's effective radius, the ``$s$-like'' symmetric
$\vec{q_1}$ mode becomes a broad band, and the ``$p$-like'' antisymmetric
$\vec{q_2}$ mode becomes a sharp proximity resonance peak, both shifted
from the peak of single scatterer, as shown in Fig.~\ref{fig1}a~\cite{hel1}. In
the limit of
$d\to 0$, $\vec{q_1}$ mode rigorously becomes an $s$-wave and $\vec{q_2}$
mode a $p$-wave. 

\begin{figure}
\includegraphics{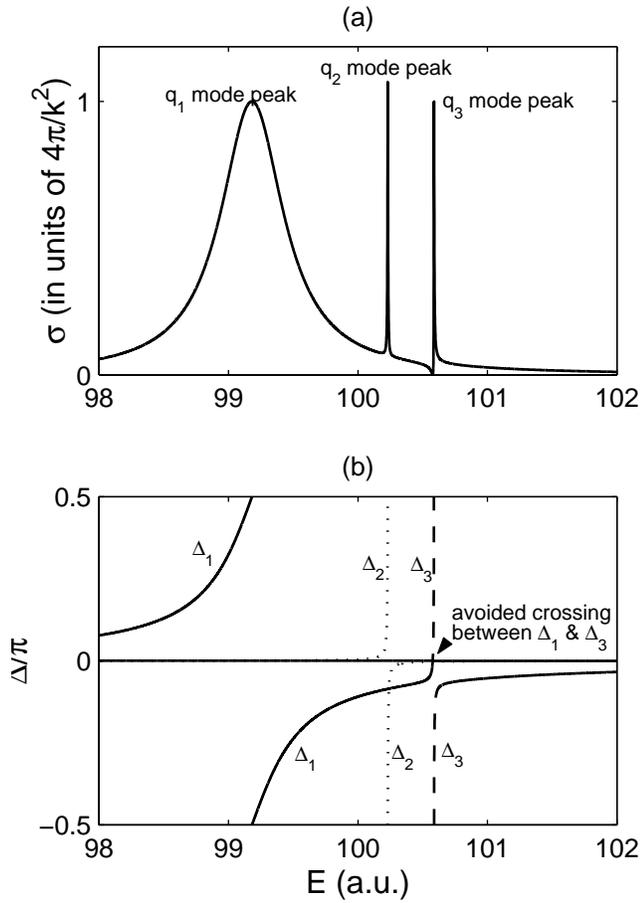}
\caption{\label{fig2} (a) Total cross section $\sigma$ and (b) phase angles $\Delta_j$ of each eigenmode as a function of energy of a system of
three identical scatterers placed on a straight line separated by
equal distance $d=0.2/k_0$, at $(-d,0,0)$, $(0,0,0)$, and 
$(d,0,0)$. Each scatterer is the same as used in
Fig.~\ref{fig1}. In (b), solid lines: $\Delta_1$, dotted lines: $\Delta_2$, dashed lines: $\Delta_3$. }
\end{figure}

The three scatterer case turns out to have a richer phenomenology. We
consider three identical scatterers positioned on a straight line
separated by equal distance: $\vec{r}_1=(-d,0,0), \vec{r}_2=(0,0,0),
\vec{r}_3=(d,0,0)$. Compared to the two scatterer case, the difference is
that the relative amplitudes in each $\vec{q_j}$ are now energy dependent. 
Nonetheless, upon reflection about $yz$ plane, there are always two
symmetric modes $\vec{q_1}$, $\vec{q_3}$, and one antisymmetric mode
$\vec{q_2}$. Fig.~\ref{fig2}b shows the phase angles as a function of energy. They
determine the positions and widths of the peaks in total cross section
(Fig.~\ref{fig2}a). The eigenmodes at $\vec{q_1}$ mode's peak
($E=99.182$) are: 
\[
\vec{q_1}=\left(\begin{array}{d}0.089\\0.106\\0.089\end{array}\right),
\mbox{ }
\vec{q_2}=\left(\begin{array}{d}1.23\\0\\-1.23\end{array}\right),
\mbox{ }
\vec{q_3}=\left(\begin{array}{d}23.9\\-47.5\\23.9\end{array}\right).
\]
We see that $\vec{q_1}$ is $s$-like, $\vec{q_2}$ $p$-like, and $\vec{q_3}$
$d$-like. Indeed numerical evidence shows that at low energies ($E<E_0$) 
$\vec{q_1}$, $\vec{q_2}$, and $\vec{q_3}$ converge to $s$-, $p$-, and
$d$-waves when $d \to 0$. The $\vec{q_1}$ mode gives most of the broad
band and the other two modes give two sharp proximity resonance peaks. 

\begin{figure}
\includegraphics{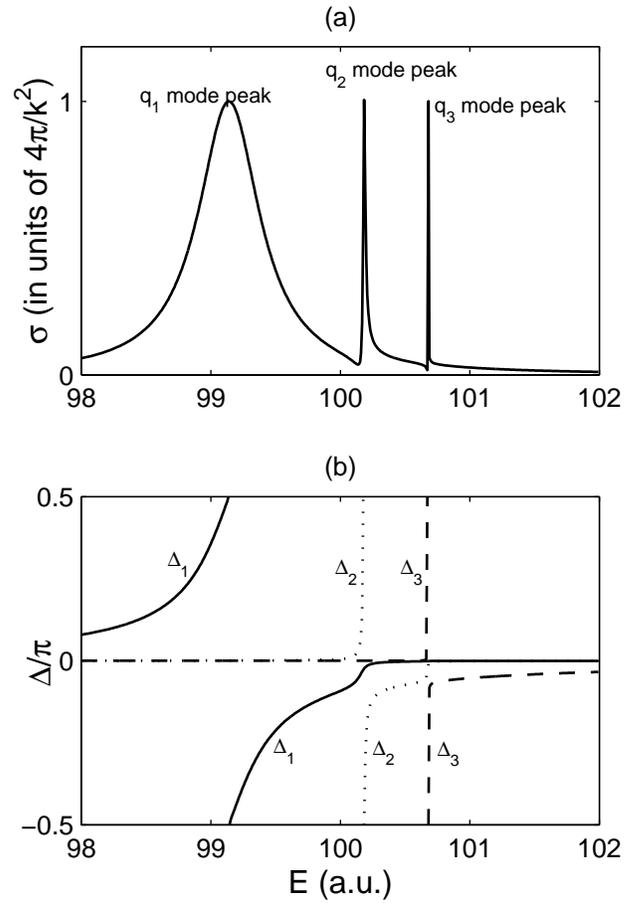}
\caption{\label{fig3}(a) Total cross section $\sigma$ and (b) phase angles
$\Delta_j$ of each
eigenmode as a function of
energy of a system of three identical scatterers placed at $(-d,0,0)$,
$(-d/3,-d/3,0)$, and $(d,0,0)$ with $d=0.2/k_0$. Each scatterer is the same
as used in Fig.~\ref{fig1}. In (b), solid lines: $\Delta_1$, dotted lines: $\Delta_2$, dashed lines: $\Delta_3$. The difference to Fig.~\ref{fig2} is that the scatterer at
the center is moved so there is no longer symmetry about the reflection
of $xz$ and $yz$ plane. }
\end{figure}

If we disregard the continuity of $\Delta_j(E)$, it is determined up to an
integral multiple of $\pi$. For reasons that will become clear in a
moment, we have drawn Fig.~\ref{fig2}b with restriction $-\pi/2 < \Delta_j
\leq \pi/2$. At $E \sim 100.578$, $\Delta_1$ and $\Delta_3$ undergo an
avoided crossing: instead of one going sharply from some $-\alpha$ to
$\pi-\alpha$ and the other remaining at $0$, $\Delta_1$ goes sharply from
$-\alpha$ to $0^-$ and $\Delta_3$ goes sharply from $0^+$ to $\pi-\alpha$.
This is really an avoided crossing of eigenvalues $\lambda_1$ and
$\lambda_3$ at $\lambda \approx 0$. Tracking $\vec{q_1}$ and $\vec{q_3}$ around the avoided crossing
region, we find $\vec{q_1}$ is $s$-like and $\vec{q_3}$ is $d$-like before
the crossing ($E<100.578$), yet $\vec{q_1}$ becomes $d$-like and $\vec{q_3}$
becomes $s$-like after the crossing ($E>100.578$).  In contrast, $\Delta_2$
rigorously crosses $\Delta_1$ and $\Delta_3$, since it has a different
symmetry. Indeed if we remove the symmetry of the system, there is always
avoided crossing whenever two phase angles tend to cross at $0$ (Fig.~\ref{fig3}b). 
The $\vec{q_3}$ mode peak exhibits a Fano lineshape (Fig.~\ref{fig2}a)~\cite{fan}. 
The dip around $E\sim100.578$ before the peak comes from the avoided
crossing at $0$.  Sin$^2 \Delta$ will exhibit a Fano lineshape if
$\Delta$ goes from some $-\alpha$ to $\pi-\alpha$, with minimum at
$\Delta=0$ and the peak still at $\Delta=\pi/2$. Therefore, there is
always a dip when two phase angles tend to cross at $0$. The deepness of
the dip depends on how narrowly they avoid crossing.

\begin{figure}
\includegraphics{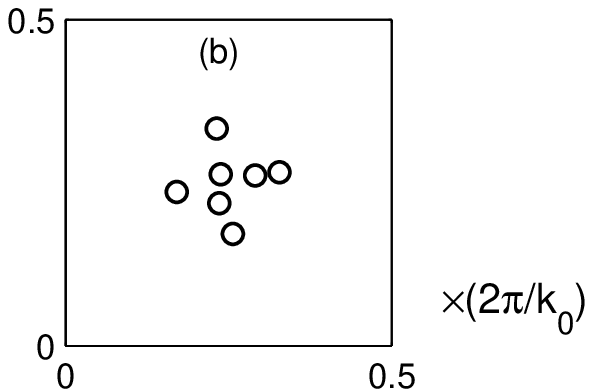}
\includegraphics{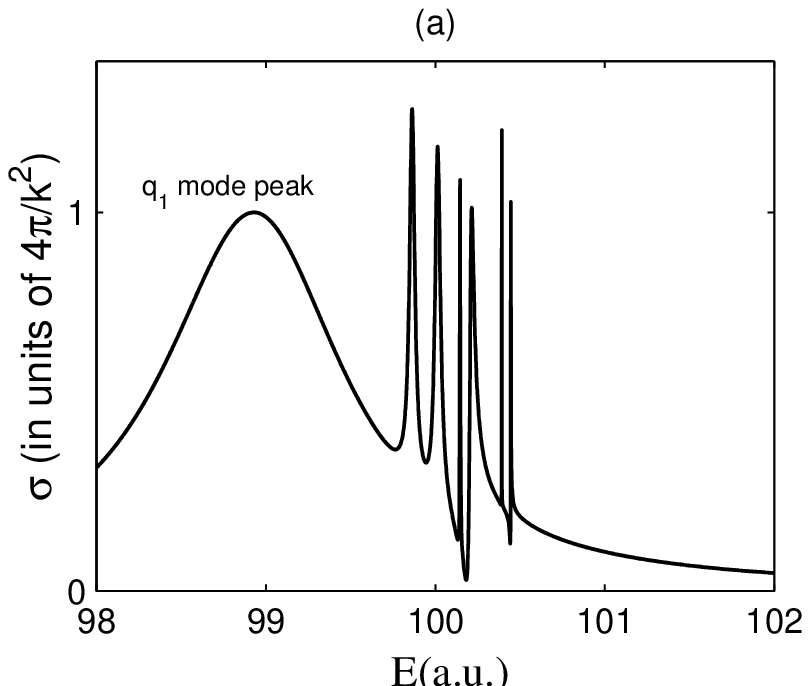}
\caption{\label{fig4} (a) Total cross section as a function of energy for a system
of seven identical scatterers randomly placed on a plane. Each scatterer is the
same as used in Fig.~\ref{fig1}. The positions of the scatterers are shown in (b).}
\end{figure}

Numerical evidence shows that for a general system of $N$ identical
scatterers confined well within each other's effective radius, there is
always a broad band with resonance energy $<E_0$ and some sharp proximity
resonance peaks based on the high energy tail of the broad band
(Fig.~\ref{fig4})~\cite{her}. At peak, the broad band corresponds to a mode (say,
$\vec{q_1}$) in which all scatterers oscillate in phase, i.e., all
$q_{i1}$'s have the same sign. The peak is broad because the waves scattered from each scatterer
add up constructively. In this sense this is an $s$-like mode, corresponding to the
superradiant state in scattering of one photon by a collection of
atoms~\cite{dic}. 
The remaining $N-1$ modes give $N-1$ sharp proximity
resonance peaks (if they are separate). They correspond to subradiant
states. The scattered waves of these modes at their respective peaks have little $s$-wave component---waves
scattered from each scatterer tend to add up
destructively. It has been reported that under certain conditions these narrow peak eigenmodes correspond to Anderson localization~\cite{rus}. $\vec{q_1}$ mode is $s$-like at low energies. As a
consequence of avoided crossing of phase angles, the $s$-like
character will shift to different eigenmodes at different energies.
Each sharp peak
corresponds to a phase angle to sharply increasing from some
$-\alpha$ ($0<\alpha<\pi/2$) on the broad band (corresponding to $s$-like mode) to
$\pi-\alpha$---back onto the broad band with avoided crossing at $0$,
exhibiting a Fano lineshape. We may not see the dip if the phase angles do
not come close enough in avoided crossing, but in principle a sharp proximity resonance peak is always unsymmetric (Fig.~\ref{fig2}a ($\vec{q_3}$ peak
only), \ref{fig3}a and \ref{fig4}a).

It is surprising to find numerically that, even for a typical
nonsymmetric system of identical scatterers, at far off-resonant energies while still maintaining $kr_{ij} \ll 1$, $y_j(\hat{r})$'s are very much like
linear combinations of spherical harmonics of same $l$, i.e., they almost
have well defined angular momenta. $y_j(\hat{r})$'s also have lowest
$l$'s possible, so they are still $s$-, $p$-, $d$-,
$\cdots$ like. At resonant energies, the eigenmodes have mixed $l$'s in
avoided crossing region.

\section{Summary}

We have shown that for a system of $N$ $s$-wave point scatterers, we can
always find $N$ eigenmodes. These eigenmodes or eigenchannels play the same role as spherical harmoics for a spherically symmetric target. The $T$
matrix of the system is of rank~$N$ and can be represented analytically. The eigenmodes are eigenvectors
corresponding to non-$0$ eigenvalues of the $T$~matrix. The position and width of resonance
peaks are determined by phase angles as a function of energy.
Each phase angle monotonically increases from $0$ to $\pi$ as energy goes
from $-\infty$ to $\infty$ if each scatterer has one Breit-Wigner type
resonance. 

For identical scatterers placed well within one wavelength of the incoming wave, the
 far off-resonant eigenmodes are always $s$-, $p$-, $d$-, $\cdots$
like with lowest possible angular momenta.                      
At resonant energies 
where phase angles have avoided crossings, the eigenmodes have components of different $l$'s mixed. There is always one broad
band originating from an $s$-like mode whose peak is shifted lower in energy, and $N-1$ sharp proximity
resonance peaks based on the high energy tail of the broad band. The broad band
peak correspond to superradiant state in scattering of light, and the
sharp peaks correspond to subradiant states. The unsymmetric Fano
lineshape of a sharp peak comes from a phase angle sharply increasing from some $-\alpha$ on the broad
band to $\pi-\alpha$ (back onto the broad band) with avoided crossing at
$0$. 

In the future, we will investigate the generalization of eigenmode
expansion to systems of multi-channel point
scatterers. We will investigate further the dependence of the
spectrum (total cross section as a function of energy)
on the spatial distribution of scatterers, and statistical
properties of the spectra of collections of a large number of randomly
distributed scatterers. We need to extend the theory to non-identical
scatterers, and to multiphoton emission of collections of atoms
(i.e. true superradiance).

\begin{acknowledgments}
We appreciate the support of the National Science Foundation, grant
CHE-0073544, and support of ITAMP (the Institute for Theoretical Atomic
and Molecular Physics) at Harvard.

\end{acknowledgments}


\end{document}